\documentclass[12pt]{article}
\usepackage{amsfonts}
\begin{document}

\def\ba{\begin{equation}}
\def\ea{\end{equation}}
\def\w{\wedge}

\begin{titlepage}
\title{{\bf  Einstein-Cartan-Dirac Theory In (1+2)-Dimensions}}
\author{ T. Dereli\footnote
{tdereli@ku.edu.tr}\\
 {\small Department of Physics, Ko\c{c} University}\\{\small 34450 Sar{\i}yer, Istanbul, Turkey} \\  \\
 N. \"{O}zdemir\footnote{nozdemir@itu.edu.tr} , \"{O}. Sert\footnote{  sertoz@itu.edu.tr}  \\
 {\small Department of Physics Engineering, Istanbul Technical University}\\
 {\small 34469 Maslak, Istanbul, Turkey  }  \\
} \vskip 1cm
\date{ }

\maketitle

\begin{abstract}

\noindent Einstein-Cartan theory is formulated in (1+2)-dimensions
using the algebra of exterior differential forms. A Dirac spinor is
coupled to gravity and the field equations are obtained by a
variational principle. The space-time torsion is found to be given
algebraically in terms of the Dirac field. Circularly symmetric,
exact solutions that collapse to  $AdS_3$ geometry in the absence of
a Dirac spinor are found.

\vskip 2cm

\end{abstract}
\end{titlepage}

\section{Introduction}

\noindent Field theories in (1+2)-dimensions often provide easy ways
to check ideas that are difficult to prove in actual
(1+3)-dimensions. However, it is not surprising to uncover other new
ideas as well that are specific to (1+2)-dimensions. Topologically
massive gravity \cite{deser} or BTZ black holes \cite{teitelboim}
are  some of the best known examples to the latter. Other aspects
and the literature may be found in Ref.\cite{carlip}. In this paper
we couple a Dirac spinor in (1+2)-dimensions to Einstein-Cartan
gravity \cite{trautman}. The field equations are derived by a
variational principle. Then a family of circularly symmetric,
rotating solutions that are asymptotically $AdS_3$ are found and
discussed.

\bigskip

\noindent We specify the space-time geometry by a triplet $ \left (
M, g, \nabla \right )$ where $M$ is a 3-dimensional differentiable
manifold equipped with a metric tensor \ba g = \eta_{ab} e^a \otimes
e^b \ea of signature  $(-++)$. $\{e^a\}$ is an orthonormal co-frame
dual to the frame vectors $\{X_a\}$, that is $e^{a}(X_b) =
\delta^{a}_{b}$. A metric compatible connection $\nabla$ can be
specified in terms of connection 1-forms $\{\omega^{a}_{\;\;b}\}$
satisfying $\omega_{ba} = -\omega_{ab}$. Then the Cartan structure
equations \ba de^a + \omega^{a}_{\;\; b} \w e^b = T^a , \ea  \ba
d\omega^{a}_{\;\; b} + \omega^{a}_{\;\; c} \w \omega^{c}_{\;\; b} =
R^{a}_{\;\; b} \ea yield the space-time torsion 2-forms $\{T^a\}$
and curvature 2-forms $\{R^{a}_{\;\;b}\}$, respectively. Here $d$
denotes the exterior derivative and $\w$ the wedge product. We fix
the orientation of space-time by choosing the volume 3-form $*1 =
e^0 \wedge e^1 \wedge e^3$ where $*$ is the Hodge star map. It is
possible to decompose the connection 1-forms in a unique way as \ba
\omega^{a}_{\;\; b} = \hat{\omega}^{a}_{\;\; b} +K^{a}_{\;\; b} \ea
where $\{\hat{\omega}^{a}_{\;\; b} \}$ are the zero-torsion
Levi-Civita connection 1-forms satisfying \ba de^a +
\hat{\omega}^{a}_{\;\; b} \w e^b = 0 \ea and $\{K^{a}_{\;\; b} \}$
are the contortion 1-forms satisfying \ba K^{a}_{\;\; b} \w e^b =
T^a .\ea The curvature 2-forms are also decomposed in a similar way:
\ba R^{a}_{\;\;b} = \hat{R}^{a}_{\;\;b} + \hat{D}K^{a}_{\;\;b} +
K^{a}_{\;\;c} \w K^{c}_{\;\;b} \ea with
$$
\hat{D}K^{a}_{\;\;b} = dK^{a}_{\;\;b} + \hat{\omega}^{a}_{\;\;c} \w
K^{c}_{\;\;b} - \hat{\omega}^{c}_{\;\;b} \w K^{a}_{\;\;c} .
$$

\medskip

\noindent The field equations of Einstein-Cartan theory of gravity
\cite{trautman} are obtained by varying the action \ba I = \int_M
\left ( {\cal{L}}_{EC} + {\cal{L}}_{M}  \right ) \ea where the
Einstein-Cartan Lagrangian density 3-form \ba {\cal{L}}_{EC} =
-\frac{1}{2 \kappa} R_{ab} \w *(e^a \w e^b) + \lambda *1 , \ea with
the gravitational constant $\kappa$ and the cosmological constant
$\lambda$, and the matter Lagrangian density 3-form ${\cal{L}}_{M}$.
We write the infinitesimal variations (up to a closed form) as \ba
\dot{\cal{L}} = \dot{e}^a \w \left ( -\frac{1}{2 \kappa} R^{bc}
*e_{abc} + \lambda
*e_a + \tau_a \right) +  \frac{1}{2} \dot{\omega}^{ab} \w \left ( -\frac{1}{\kappa}*e_{abc} T^c +
\Sigma_{ab} \right )  \ea where the variations of the matter
Lagrangian yield the stress-energy 2-forms  \ba \tau_a =
\frac{\partial {\cal{L}_{M}}}{\partial e^a} = T_{ab} *e^b \ea and
the angular momentum 2-forms \ba \Sigma_{ab} = \frac{\partial
{\cal{L}_{M}}}{\partial \omega^{ab}} = S_{ab,c} *e^c . \ea Therefore
the Einstein-Cartan  field equations are given either as \ba
\frac{1}{2} R^{bc}
*e_{abc} - \kappa \lambda
*e_a  = \kappa \tau_a , \ea
 \ba *e_{abc} T^c = \kappa \Sigma_{ab} \ea
or equivalently as  \ba R_{ab} =
 \kappa \lambda ( e_a \w e_b) +  \kappa *e_{abc} \tau^c \label{einstein} , \ea \ba  T_a = \frac{1}{2} \kappa *e_{abc} \Sigma^{bc} .
 \label{cartan}  \ea

\section{Einstein-Cartan-Dirac Field Equations}

\noindent Let us consider  a Dirac spinor field \ba \psi = \left (
\begin{array}{c} \psi_1 \\ \psi_2 \end{array}
\right ) \ea and the conjugate spinor field \ba \bar{\psi} =
\psi^{\dagger} \gamma_0 = \left ( -\psi^{*}_2  \; \; \psi^{*}_1
\right ) \ea where $\psi_1$ and $\psi_2$ are complex, odd Grassmann
valued functions.  We use a real (i.e. Majorana) realization of the
gamma matrices $\{\gamma_a\}$ given explicitly as \ba \gamma_0 =
\left (
\begin{array}{cc} 0 & 1\\ -1 & 0 \end{array} \right ) \; ,
\;\gamma_1 = \left (  \begin{array}{cc} 0 & 1\\ 1 & 0
\end{array} \right ) \; , \;\gamma_2 = \left (  \begin{array}{cc} 1 & 0\\0& -1  \end{array}
\right ) . \ea The exterior covariant derivative of the spinor
fields are defined to be \ba \nabla \psi = d \psi + \frac{1}{2}
\omega^{ab} \sigma_{ab} \psi \quad , \quad \overline{\nabla \psi} =
d \bar{\psi} -\frac{1}{2} \omega^{ab} \bar{\psi} \sigma_{ab} \ea
with \ba \sigma_{ab} = \frac{1}{4} \left [ \gamma_{a} ,\gamma_{b}
\right ] = \frac{1}{2}
*e_{abc} \gamma^c . \ea
We set $\gamma = \gamma_{a}  e^{a} .$

\medskip

\noindent Next let us introduce the (Hermitian)  Dirac Lagrangian
density 3-form \ba {\cal{L}}_D = \frac{i}{2} \left ( \bar{\psi}
*\gamma \w \nabla \psi - \overline{\nabla \psi} \w *\gamma \psi
\right ) + i m \bar{\psi} \psi
*1 . \ea
Its infinitesimal variations are found to be (up to a closed form)
\begin{eqnarray} {\dot{\cal{L}}}_D  &=& \dot{e}^a \w \left \{ \frac{i}{2}
*e^{b}_{\; \; a} \w ( \bar{\psi} \gamma_b \nabla \psi + \overline{\nabla \psi} \gamma_{b} \psi) + i m \bar{\psi} \psi *e_a \right
\} \nonumber \\ & &  + \frac{1}{2} \dot{\omega}^{ab} \w \left \{
\frac{i}{2} \bar{\psi} (
*\gamma \sigma_{ab} + \sigma_{ab} *\gamma ) \psi
 \right \} \nonumber \\
& & +  i \dot{\bar{\psi}} \left \{ *\gamma \w \nabla \psi +
\frac{1}{2}
*e^{a}_{\;\; b} \w T^b \gamma_a \psi + m * \psi   \right \} \nonumber \\
& & - i  \left \{ \overline{\nabla \psi} \w *\gamma  - \frac{1}{2}
*e^{a}_{\;\;b} \w T^b  \bar{\psi} \gamma_a - m * \bar{\psi}   \right \} \dot{\psi} .
\end{eqnarray}
From the above expression we identify the stress-energy 2-forms \ba
\tau_a = \frac{i}{2} *e^{b}_{\;\;a} \left ( \bar{\psi} \gamma_b
\nabla \psi + \overline{\nabla \psi} \gamma_b \psi \right ) +  i m
\bar{\psi} \psi *e_a \ea and the angular momentum 2-forms \ba
\Sigma_{ab} = \frac{i}{2} \left ( \bar{\psi}  *\gamma \sigma_{ab}
\psi + \bar{\psi} \sigma_{ab} *\gamma \psi  \right ) .\ea
Substituting these into the Einstein-Cartan equations
(\ref{einstein}) and {\ref{cartan}) we obtain
\begin{eqnarray}
R_{ab} &=& \kappa \lambda e_a \w e_b + i m \kappa \bar{\psi} \psi
e_a \w e_b \nonumber \\
& & + i \frac{\kappa}{2} e_a \w ( \bar{\psi} \gamma_b \nabla \psi )
- i \frac{\kappa}{2} e_b \w ( \bar{\psi} \gamma_a \nabla \psi )
\nonumber \\ & & - i \frac{\kappa}{2} e_b \w ( \overline{\nabla
\psi} \gamma_a \psi ) +  i \frac{\kappa}{2} e_a \w ( \bar{\nabla
 \psi} \gamma_b \psi ) , \label{curvature}
\end{eqnarray}
\ba T_a = i \frac{\kappa}{2} \bar{\psi} \psi *e_a \label{torsion}
\ea and the Dirac equation \ba
*\gamma \w \nabla \psi + m \psi *1 = 0 . \label{dirac}
\ea We note that with the torsion 2-forms (\ref{torsion}),
$*e^{a}_{\;\;b} \w T^b = 0$  identically and the Dirac equation
 simplifies to (\ref{dirac}).

\medskip

\noindent These field equations can be re-written in terms of the
Levi-Civita connection only. To see this, we solve (\ref{torsion})
for the contortion 1-forms \ba K_{ab} = -i \frac{\kappa}{4}
\bar{\psi} \psi
*(e_a \w e_b)  \ea and substitute these into (\ref{curvature}) which
simplify to \ba \hat{R}_{ab} = \kappa \lambda e_a \w e_b + *e_{abc}
{\hat{\tau}}^c -i \frac{\kappa}{4} d( \bar{\psi} \psi ) \w *(e_a \w
e_b) + \frac{3 \kappa^2}{16} (\bar{\psi} \psi )^2 e_a \w e_b .
\label{einstein2}\ea The Dirac equation (\ref{dirac}) similarly
simplifies to \ba
*\gamma \w \bar{\nabla} \psi + m \psi *1 -i \frac{3 \kappa}{8} (\bar{\psi} \psi ) \psi *1  =
0 \label{dirac2} \ea

\medskip

\noindent It is interesting to note that the Einstein-Dirac
equations (\ref{einstein2}) and (\ref{dirac2}) can be obtained from
an action by  zero-torsion constrained variations using the method
of Lagrange multipliers \cite{dereli1988}. To this end, we consider
a modified  Dirac Lagrangian density 3-form \ba {\cal{L}}_D^{\prime}
= \frac{i}{2} \left ( \bar{\psi}
*\gamma \w \hat{\nabla} \psi - \overline{{\hat{\nabla}} \psi} \w *\gamma \psi
\right ) + i m \bar{\psi} \psi *1 + \frac{3 \kappa^2}{16} (
\bar{\psi} \psi )^2
*1  \ea
together with the constraint term \ba {\cal{L}}_{constraint} = (
de^a + \omega^{a}_{\;\;b} \w e^b ) \w \lambda_a \ea where
$\{\lambda_a\}$ are the Lagrange multiplier 1-forms. The variation
of the total action
$$
I = \int_M \left ( {\cal{L}}_{EC} + {\cal{L}}_D^{\prime} +
{\cal{L}}_{constraint} \right )
$$
with respect to the Lagrange multipliers imposes the constraint that
the connection 1-forms are Levi-Civita. Then we solve the connection
variation equations under this constraint for the Lagrange
multiplier 1-forms as
$$
\lambda_a = -\frac{i}{4} \kappa \bar{\psi} \psi e_a .
$$
Substituting these in the remaining Einstein and Dirac equations
give precisely (\ref{einstein2}) and (\ref{dirac2}).

\section{A Circularly Symmetric Solution}

\noindent We consider the metric  \ba g = - f^2(r) dt^2 + h^2(r)
dr^2 + r^2 ( d \phi + a(r) dt )^2 \ea in plane polar coordinates
$(t, r, \phi)$ \cite{dereli2000}. The following choice of
orthonormal basis 1-forms \ba e^0 = f(r) dt \quad , \quad e^1 = h(r)
dr \quad , \quad e^2 = r ( d \pi + a(r) dt ) , \ea leads to the
Levi-Civita connection 1-forms \ba \hat{\omega}^{0}_{\; \;1} =
\alpha e^0 - \frac{\beta}{2} e^2 \; , \; \hat{\omega}^{0}_{\; \; 2}
= -\frac{\beta}{2} e^1 \; , \; \hat{\omega}^{1}_{\; \; 2} = -\gamma
e^2 -\frac{\beta}{2} e^0 \ea where we defined \ba \alpha =
\frac{f^{\prime}}{f h } \quad , \quad \beta = \frac{r a^{\prime}}{f
h } \quad , \quad \gamma = \frac{1}{r h} \ea with ${ }^\prime$
denoting the derivative $\frac{d}{dr}$. On the other hand, assuming
$ i \frac{\kappa}{2} \bar{\psi} \psi = \tau(r)$, we calculate the
contortion 1-forms \ba K^{0}_{\; \;1} = \frac{\tau}{2} e^2 \; , \;
K^{0}_{\; \; 2} = -\frac{\tau}{2} e^1 \; , \; K^{1}_{\; \; 2} =
-\frac{\tau}{2} e^0 . \ea As a first step towards a solution, we
take a Dirac spinor that depends only on $r$ and work out
(\ref{dirac}) in components as follows:
\begin{eqnarray}
\psi^{\prime}_{1} + \frac{h}{2} ( \alpha + \gamma ) \psi_1 +
\frac{h}{4} ( \beta + 3 \tau + 4m) \psi_2 &=& 0 , \nonumber \\
\psi^{\prime}_{2} + \frac{h}{2} ( \alpha + \gamma ) \psi_2 +
\frac{h}{4} ( \beta + 3 \tau + 4m) \psi_1 &=& 0 .
\end{eqnarray}
We take the combinations $\psi_{+} = \psi_1 + \psi_2$ and $\psi_{-}
= \psi_1 - \psi_2$ and write a decoupled system of equations
\begin{eqnarray}
\psi^{\prime}_{+} + (k_1 + k_2 ) \psi_{+} &=& 0 \nonumber \\
\psi^{\prime}_{-} + (k_1 - k_2 ) \psi_{-} &=& 0
\end{eqnarray}
where we set
$$
k_1 = \frac{h}{2} ( \alpha + \gamma )  \quad , \quad k_2 =
\frac{h}{4} ( \beta + 3 \tau + 4m).
$$
The formal solution to these equations are given by
\begin{eqnarray}
\psi_1 &=& e^{-\int^r k_1 dr} \left ( \xi_1 e^{-\int^r k_2 dr} +
\xi_2 e^{\int^r k_2 dr} \right )   \nonumber \\ \psi_2 &=&
e^{-\int^r k_1 dr} \left ( \xi_1 e^{\int^r k_2 dr} - \xi_2
e^{-\int^r k_2 dr} \right )
\end{eqnarray}
where $\xi_1$ and $\xi_2$ are complex, odd Grassmann valued
constants. It can easily be verified \ba \tau(r) =  i \kappa (
\xi^{*}_2 \xi_1 - \xi^{*}_1 \xi_2 ) e^{-2 \int^r k_1 dr} . \ea

\medskip

\noindent We next work out the Einstein field equations
(\ref{einstein}) which after simplifications reduce to the following
system of coupled first order differential equations:
\begin{eqnarray}
\frac{\beta^{\prime}}{2 h} + \beta \gamma & = & -
\frac{\tau^{\prime}}{2 h} - \tau \alpha \nonumber \\
\frac{\gamma^{\prime}}{h} + \frac{\beta^2}{4} + \gamma^2 + \kappa
\lambda &=& \frac{3 \tau^2}{4} + \frac{\beta \tau}{2}
\nonumber \\
\frac{\alpha^{\prime}}{h} - \frac{3\beta^2}{4} + \alpha^2 + \kappa
\lambda &=& \frac{3 \tau^2}{4} - \frac{\beta \tau}{2}
\nonumber \\
\frac{\beta^2}{4} + \alpha \gamma + \kappa \lambda & =& -\frac{3
\tau^2}{4} - 2 m \tau \nonumber \\
 \frac{\beta^{\prime}}{2 h} + \beta \gamma & = &
\frac{\tau^{\prime}}{2 h} + \tau \gamma  . \label{equations}
\end{eqnarray}
At this point we choose $$ \kappa \lambda = - \frac{1}{l^2} < 0 $$
and restrict our attention to those cases for which
$$ \gamma = \alpha = \frac{1}{r h} \quad , \quad \tau = \beta =
\frac{\beta_0}{r^2} .
$$
We can then integrate (\ref{equations}) for the metric functions \ba
f(r) = \frac{r}{l} \quad , \quad h(r) = \frac{l}{r \sqrt{ 1 -
\frac{2 m \beta_0 l^2}{r^2} - \frac{\beta^2_0 l^4}{r^4}}},\ea and
\ba a(r) = \frac{1}{2l} \arcsin \left ( \frac{m}{\sqrt{m^2 +
\frac{1}{l^2}}} \right ) - \frac{1}{2l} \arcsin \left ( \frac{m +
\frac{\beta_0}{r^2}}{\sqrt{m^2 + \frac{1}{l^2}}} \right ) . \ea It
now remains to integrate for the Dirac spinor and we find \ba
\psi_{\pm} = \frac{1}{r} \left |r^2 - m \beta_0 l^2 + \sqrt{r^4 - 2
m \beta_0 l^2 r^2 - \beta^2_0 l^4 }\right |^{\pm \frac{ml}{2}}
e^{\mp \frac{1}{2 \beta_0} \arcsin \left ( \frac{m +
\frac{\beta_0}{r^2}}{\sqrt{m^2 + \frac{1}{l^2}}} \right ) }
\xi_{\pm} \ea where $\xi_{\pm} = \xi_1 \pm \xi_2 .$

\medskip

\noindent In order to understand the physical meaning of this
solution we write down the metric \ba g = - \frac{r^2}{l^2} dt^2 +
\frac{l^2 dr^2}{r^2 \left ( 1 -\frac{2 m \beta_0 l^2}{r^2} -
\frac{\beta_0^2 l^2}{r^4}\right ) } + r^2 (d\phi + a(r) dt)^2 . \ea
We see that in the absence of a Dirac spinor ($\beta_0 = 0$) the
above metric collapses to the $AdS_3$ metric \ba g_0 =
-\frac{r^2}{l^2} dt^2 + \frac{l^2}{r^2} dr^2 + r^2 d\phi^2 . \ea
Even when $\beta_0 \neq 0$, the metric $g \rightarrow g_0$
asymptotically as $r \rightarrow \infty $. We note  a metric
singularity at \ba r_c = \left \{
\begin{array}{lr} l \sqrt{m \beta_0 + \beta_0 \sqrt{m^2 +
\frac{1}{l^ 2}}} \quad & , \quad \beta_0 > 0 \\
l \sqrt{|\beta_0| \sqrt{m^2 + \frac{1}{l^ 2}}- m |\beta_0|} \quad &
, \quad \beta_0 < 0 \end{array} \right . \ea We further calculate
the curvature scalar \ba {\cal{R}} = -\frac{6}{l^2} + \frac{4 m
\beta_0}{r^2} \ea and the quadratic curvature invariant \ba R_{ab}
\w *R^{ab} =  \frac{6}{l^4} -\frac{8 m \beta_0}{l^2 r^2} +
\frac{\beta_0^2 (8m^2-\frac{4}{l^2})}{r^4} +  \frac{16 m
\beta_0^2}{r^6} + \frac{8 \beta_0^4}{r^8} . \ea  The curvature
invariants have an essential singularity at $r = 0$. We also check
(See Ref.\cite{dereli2000}) the quasi-local angular momentum \ba
J(r) = \frac{r^3}{f(r) h(r)} \frac{da}{dr} = \beta_0 , \ea the
quasi-local energy \ba E(r) = \frac{1}{h_0(r)} - \frac{1}{h(r)} =
\frac{r}{l} - \sqrt{\frac{r^2}{l^2} - 2 m \beta_0 -
\frac{\beta_0^2}{r^2}} \simeq \frac{\beta_0 m l}{r} ,\ea and the
quasi-local mass \begin{eqnarray} M(r) & =& 2 f(r) E(r) = J(r)a(r)
\nonumber \\ &=& 2\frac{l^2}{r^2} - 2\frac{l^2}{r^2} \sqrt{1
-\frac{2 m \beta_0 l^2}{r^2} - \frac{\beta_0^2 l^2}{r^4}} \nonumber
\\ & & + \frac{1}{2l} \arcsin \left ( \frac{m}{\sqrt{m^2 +
\frac{1}{l^2}}} \right ) - \frac{1}{2l} \arcsin \left ( \frac{m +
\frac{\beta_0}{r^2}}{\sqrt{m^2 + \frac{1}{l^2}}} \right ) \\
&\simeq& 2 m \beta_0  . \nonumber
\end{eqnarray}

\newpage

\section{Conclusion}

\noindent We have formulated the Einstein-Cartan-Dirac theory in
(1+2)-dimensions using the algebra of exterior differential forms.
We coupled a Dirac spinor to Einstein-Cartan gravity and obtained
the field equations by a variational principle. The space-time
torsion is given algebraically in terms of the quadratic spinor
invariant. We then looked for rotating, circularly symmetric
solutions. We found a particular class of solutions that possesses
an essential curvature singularity at the origin $r=0$ that is
hidden behind an event horizon at some finite distance $r=r_c$ away
from the origin. The mass and the spin of this configuration can be
identified. Thus the space-time geometry we found exhibits all the
essential features of a black hole and we find it interesting that
in the absence of the Dirac spinor field collapses to the regular
$AdS_3$ geometry.

\bigskip

\noindent {\bf {\Large Acknowledgement}}

\medskip

\noindent TD is supported in part by the Turkish Academy of Sciences
(T\"{U}BA).

\end{document}